\documentclass[floatfix,intlimits,pra,showkeys,showpacs,twocolumn]{revtex4-1}

\usepackage{amsfonts,amsmath,amssymb}
\usepackage{dcolumn}
\usepackage[dvips]{graphicx}
\usepackage{hyperref}

\begin{document}

\title{A new concept multi-stage Zeeman decelerator}

\author{Theo Cremers}
\author{Simon Chefdeville}
\author{Niek Janssen}
\author{Edwin Sweers}
\author{Sven Koot}
\author{Peter Claus}
\author{Sebastiaan Y.T. van de Meerakker}
\email{basvdm@science.ru.nl}

\affiliation{Radboud University, Institute for Molecules and Materials, Heijendaalseweg 135, 6525 AJ Nijmegen, the Netherlands}

\date{\today}

\begin{abstract}
We present a new concept for a multi-stage Zeeman decelerator that is optimized particularly for applications in molecular beam scattering experiments. The decelerator consists of a series of alternating hexapoles and solenoids, that effectively decouple the transverse focusing and longitudinal deceleration properties of the decelerator. It can be operated in a deceleration and acceleration mode, as well as in a hybrid mode that makes it possible to guide a particle beam through the decelerator at constant speed. The deceleration features phase stability, with a relatively large six-dimensional phase-space acceptance. The separated focusing and deceleration elements result in an unequal partitioning of this acceptance between the longitudinal and transverse directions. This is ideal in scattering experiments, which typically benefit from a large longitudinal acceptance combined with narrow transverse distributions. We demonstrate the successful experimental implementation of this concept using a Zeeman decelerator consisting of an array of 25 hexapoles and 24 solenoids. The performance of the decelerator in acceleration, deceleration and guiding modes is characterized using beams of metastable Helium ($^3S$) atoms. Up to 60\% of the kinetic energy was removed for He atoms that have an initial velocity of 520 m/s. The hexapoles consist of permanent magnets, whereas the solenoids are produced from a single hollow copper capillary through which cooling liquid is passed. The solenoid design allows for excellent thermal properties, and enables the use of readily available and cheap electronics components to pulse high currents through the solenoids. The Zeeman decelerator demonstrated here is mechanically easy to build, can be operated with cost-effective electronics, and can run at repetition rates up to 10~Hz.
\end{abstract}

\maketitle

\section{Introduction}\label{sec:intro}
In the last two decades, tremendous progress has been made in manipulating the motion of molecules in a molecular beam. Using methods that are inspired by concepts from charged particle accelerator physics, complete control over the velocity of molecules in a beam can be achieved. In particular, Stark and Zeeman decelerators have been developed to control the motion of molecules that possess an electric and magnetic dipole moment using time-varying electric and magnetic fields, respectively. Since the first experimental demonstration of Stark deceleration in 1998 \cite{Bethlem:PRL83:1558}, several decelerators ranging in size and complexity have been constructed \cite{Meerakker:CR112:4828,Narevicius:ChemRev112:4879,Hogan:PCCP13:18705}. Applications of these controlled molecular beams are found in high-resolution spectroscopy, the trapping of molecules at low temperature, and advanced scattering experiments that exploit the unprecedented state-purity and/or velocity control of the packets of molecules emerging from the decelerator \cite{Carr:NJP11:055049,Bell:MolPhys107:99,Jankunas:ARPC66:241,Stuhl:ARPC65:501,Brouard:CSR43:7279,Krems:ColdMolecules}.

Essential in any experiment that uses a Stark or Zeeman decelerator is a high particle density of the decelerated packet. For this, it is imperative that the molecules are decelerated with minimal losses, i.e., molecules within a certain volume in six-dimensional (6D) phase-space should be kept together throughout the deceleration process \cite{Bethlem:PRL84:5744}. It is a formidable challenge, however, to engineer decelerators that exhibit this so-called phase stability. The problem lies in the intrinsic field geometries that are used to manipulate the beam. In a multi-stage Zeeman (Stark) decelerator a series of solenoids (high-voltage electrodes) yields the deceleration force as well as the transverse focusing force. This can result in a strong coupling between the longitudinal (forward) and transverse oscillatory motions; parametric amplification of the molecular trajectories can occur, leading to losses of particle density \cite{Meerakker:PRA73:023401,Sawyer:EPJD48:197}.

For Stark decelerators, the occurrence of instabilities can be avoided without changing the electrode design. By operating the decelerator in the so-called $s=3$ mode \cite{Meerakker:PRA71:053409}, in which only one third of the electrode pairs are used for deceleration while the remaining pairs are used for transverse focusing, instabilities are effectively eliminated \cite{Meerakker:PRA73:023401,Scharfenberg:PRA79:023410}. The high particle densities afforded by this method have recently enabled a number of high-resolution crossed beam scattering experiments, for instance \cite{Gilijamse:Science313:1617,Kirste:Sience338:1060,Zastrow:NatChem6:216,Vogels:SCIENCE350:787}. For multi-stage Zeeman decelerators, several advanced switching protocols have been proposed and tested to mitigate losses. Wiederkehr \emph{et al.} extensively investigated phase stability in a Zeeman decelerator, particularly including the role of the nonzero rise and fall times of the current pulses, as well as the influence of the operation phase angle \cite{Wiederkehr:JCP135:214202,Wiederkehr:PRA82:043428}. Evolutionary algorithms were developed to optimize the switching pulse sequence, significantly increasing the number of particles that exit from the decelerator. Furthermore, inspired by the $s=3$ mode of a Stark decelerator, alternative strategies for solenoid arrangements were investigated numerically \cite{Wiederkehr:PRA82:043428}. Dulitz \emph{et al.} developed a model for the overall 6D phase-space acceptance of a Zeeman decelerator, from which optimal parameter sets can be derived to operate the decelerator at minimum loss \cite{Dulitz:PRA91:013409}. Dulitz \emph{et al.} also proposed and implemented schemes to improve the transverse focusing properties of a Zeeman decelerator by applying reversed current pulses to selected solenoids \cite{Dulitz:JCP140:104201}. Yet, despite the substantial improvements these methods can offer, the phase-stable operation of a multi-stage Zeeman decelerator over a large range of velocities remains challenging.

Recently, a very elegant approach emerged that can be used to overcome these intrinsic limitations of multi-stage decelerators. So-called traveling wave decelerators employ spatially moving electrostatic or magnetic traps to confine part of the molecular beam in one or multiple wells that start traveling at the speed of the molecular beam pulse and are subsequently gradually slowed down. In this approach the molecules are confined in genuine potential wells, and stay confined in these wells until the final velocity is reached. Consequently, these decelerators are inherently phase stable, and no losses occur due to couplings of motions during the deceleration process. The acceptances are almost equal in both the longitudinal and transverse directions, which appears to be particularly advantageous for experiments that are designed to spatially trap the molecules at the end of the decelerator. Both traveling wave Stark \cite{Osterwalder:PRA81:051401,vandenBerg:JMS300:201422} and Zeeman \cite{Trimeche:EPJD65:263,Lavert-Ofir:NJP13:103030,Lavert-Ofir:PCCP13:18948,Akerman:NJP17:065015} decelerators have been successfully demonstrated. Recently, first experiments in which the decelerated molecules are subsequently loaded into static traps have been conducted \cite{Quintero:PRL110:133003,Jansen:PRA88:043424}.

These traveling wave decelerators typically feature a large overall 6D acceptance. This acceptance is almost equally partitioned between the longitudinal and both transverse directions. For high-resolution scattering experiments, however, there are rather different requirements for the beam than for trapping. Certainly, phase-stable operation of the decelerator---and the resulting production of molecular packets with high number densities---is essential. In addition, tunability over a wide range of final velocities is important, but the ability to reach very low final velocities approaching zero meters per second is often inconsequential. More important is the shape of the emerging packet in phase-space, i.e., the spatial and velocity distributions in both the longitudinal and transverse directions.

Ideally, for scattering experiments the longitudinal acceptance of the decelerator should be relatively large, whereas it should be small in the transverse directions. A broad longitudinal distribution---in the order of a few tens of mm spatially and 10--20 m/s in velocity---is typically required to yield sufficiently long interaction times with the target beam or sample, and to ensure the capture of a significant part of the molecular beam pulse that is available for scattering. In addition, a large longitudinal velocity acceptance allows for the application of advanced phase-space manipulation techniques such as bunch compression and longitudinal cooling to further improve the resolution of the experiment \cite{Crompvoets:PRL89:093004}. By contrast, much narrower distributions are desired in the transverse directions. Here, the spatial diameter of the beam should be matched to the size of the target beam and the detection volume; typically a diameter of several mm is sufficient. Finally, the transverse velocity distribution should be narrow to minimize the divergence of the beam. These desiderata on beam distributions are unfortunately not met by traveling wave decelerators, where the resulting longitudinal (spatial) distributions are smaller and the transverse distributions are larger than what may be considered ideal for scattering experiments.

Here, we describe a new concept for a multi-stage Zeeman decelerator that is optimized for applications in scattering experiments. The decelerator consists of an array of alternating magnetic hexapoles and solenoids, used to effectively decouple the longitudinal and transverse motions of the molecules inside the decelerator. We analyze in detail the performance of the decelerator using numerical trajectory calculations, and we will show that the decelerator exhibits phase stability, with a spatial and velocity acceptance that is much larger in the longitudinal than in the transverse directions. We show that the decelerator is able to both decelerate and accelerate, as well as to guide a packet of molecules through the decelerator at constant speed. We present the successful experimental implementation of the concept, using a multi-stage Zeeman decelerator consisting of 24 solenoids and 25 hexapoles. The performance of the decelerator in acceleration, deceleration and guiding modes is characterized using a beam of metastable helium atoms.

In the decelerator presented here, we use copper capillary material in a new type of solenoid that allows for direct contact of the solenoid material with cooling liquid. The solenoid is easily placed inside vacuum, it offers excellent thermal properties and it allows for the use of low-voltage electronic components that are readily available and cost effective. Together, this results in a multi-stage Zeeman decelerator that is relatively easy and cheap to build, and that can be operated at repetition rates up to 10 Hz.

This paper is organized as follows. In section \ref{sec:concept} we first describe the concept of the multi-stage Zeeman decelerator and characterize its inherent performance with numerical simulations. For this, we use decelerators of arbitrary length and the NH ($X\,^3 \Sigma^-$) radical as an example, as this molecule is one of our target molecules for future scattering experiments. In the simulations, we use the field geometry as induced by the experimentally proven solenoid used in the Zeeman decelerator at ETH Z{\"u}rich \cite{Wiederkehr:JCP135:214202}. In section \ref{sec:experiment}, we describe in a proof-of-principle experiment the successful implementation of the concept. Here, we use metastable helium atoms, as this species can be decelerated significantly using the relatively short decelerator presently available.

\section{Zeeman decelerator concept and design} \label{sec:concept}
\begin{figure}[!htb]
    \centering
    \resizebox{1.0\linewidth}{!}
    {\includegraphics{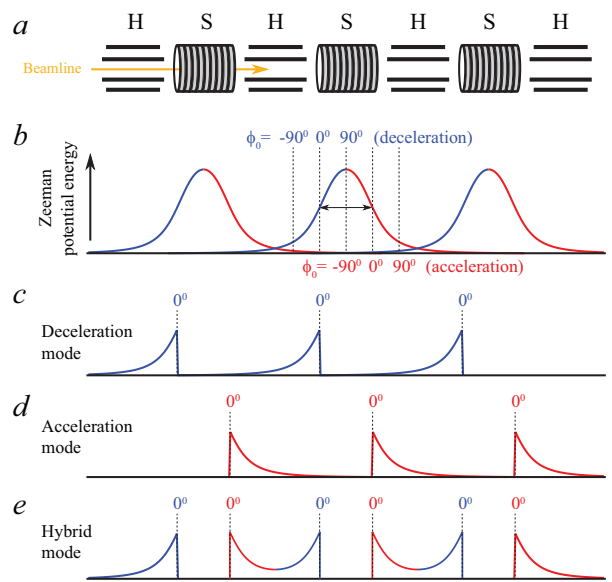}}
    \caption{(\emph{a}) Schematic representation of the array of alternating hexapoles (H) and solenoids (S). (\emph{b}) Zeeman potential energy experienced by a low-field seeking molecule traveling along the beam axis for activated solenoids. The decelerator can be operated in a deceleration mode (\emph{c}), an acceleration mode (\emph{d}), and a hybrid mode (\emph{e}). For each mode, the Zeeman energy experienced by the synchronous molecule is illustrated for $\phi_0=0^{\circ}$. Blue and red curves indicate the part of the potentials that decelerate and accelerate low-field seeking particles, respectively.}
    \label{fig:mode-explain}
\end{figure}

\begin{figure}[!htb]
    \centering
    \resizebox{1.0\linewidth}{!}
    {\includegraphics{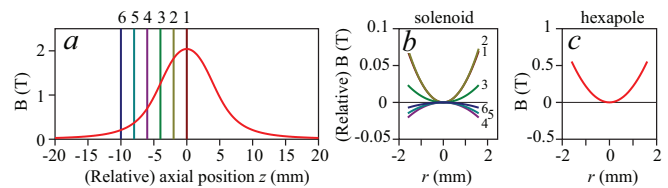}}
    \caption{(\emph{a}) On-axis magnetic field strength produced by a solenoid. (\emph{b}) Radial field induced by the solenoid at various axial positions as indicated by the colored and numbered lines in panel \emph{a}. (\emph{c}) Radial field inside a hexapole.}
    \label{fig:coilhexafield}
\end{figure}

The multi-stage Zeeman decelerator we propose consists of a series of alternating hexapoles and solenoids, as is shown schematically in Figure \ref{fig:mode-explain}. The length of the hexapoles and solenoids are almost identical. To simulate the magnetic field generated by the solenoids, we choose parameters that are similar to the ones used in the experiments by Wiederkehr \emph{et al.} \cite{Wiederkehr:JCP135:214202}. We assume a solenoid with a length of 7.5~mm, an inner and outer diameter of 7~and 11~mm, respectively, through which we run maximum currents of 300~A. Furthermore, we set the inner diameter to 3~mm for molecules to pass through. These solenoids can, for instance, be produced by winding enameled wire in multiple layers, and the current through these solenoids can be switched using commercially available high-current switches. With these levels of current, this solenoid can create a magnetic field strength on the molecular beam axis as shown in Figure \ref{fig:coilhexafield}\emph{a}; the radial profiles of the field strength at a few positions $z$ along the beam axis are shown in panel \emph{b}. It is shown that the solenoid creates a concave field distribution near the center of the solenoid, whereas a mildly convex shape is produced outside the solenoid.

The hexapoles have a length of 8.0~mm, are separated by a distance $D=4$~mm from the solenoids, and produce a magnetic field that is zero on the molecular beam axis but that increases quadratically as a function of the radial off-axis position $r$ (see Figure \ref{fig:coilhexafield}\emph{c}). We assume that the maximum magnetic field strength amounts to 0.5~T at a radial distance $r=1.5$~mm from the beam axis. Such magnetic field strengths are readily produced by arrangements of current carrying wires, permanent magnets \cite{Watanabe:EPJD38:219,Osterwalder:EPJ-TI2:10}, or a combination of both \cite{Poel:NJP17:055012}.

The key idea behind this Zeeman decelerator concept is to effectively decouple the longitudinal and transverse motions of the molecules inside the decelerator. The fields generated by the solenoids are used to decelerate or accelerate the beam, but their mild transverse focusing and defocusing forces are almost negligible compared to the strong focusing effects of the hexapoles. These hexapoles, in turn, hardly contribute to the longitudinal deceleration forces. As we will discuss more quantitatively in the next sections, this stabilizes molecular trajectories and results in phase stability.

Decelerators in which dedicated and spatially separated elements are used for transverse focusing and longitudinal deceleration have been considered before \cite{Kalnins:RSI73:2557,Sawyer:EPJD48:197}. In charged particle accelerators, such separation is common practice, and the detrimental effects of elements that affect simultaneously the longitudinal and transverse particle motions are well known \cite{Lee:AccPhys:2004}. The insertion of focusing elements between the mechanically coupled deceleration electrodes in a Stark decelerator appears technically impractical, however. By contrast, the relatively open structure of individually connected solenoids in a Zeeman decelerator allows for the easy addition of focusing elements. In addition, magnetic fields generated by adjacent elements are additive; shielding effects of nearby electrodes that are a common problem when designing electric field geometries do not occur.

The insertion of hexapoles further opens up the possibility to operate the Zeeman decelerator in three distinct modes that allow for either deceleration, acceleration, or guiding the molecular packet through the decelerator at constant speed. These operation modes are schematically illustrated in the lower half of Figure \ref{fig:mode-explain}. In the description of the decelerator, we use the concepts of an equilibrium phase angle $\phi_0$ and a synchronous molecule from the conventions used to describe Stark decelerators \cite{Bethlem:PRL83:1558,Bethlem:PRA65:053416}. The definition of $\phi_0$ in each of the modes is illustrated in Figure \ref{fig:mode-explain}, where zero degrees is defined as the relative position along the beam axis where the magnetic field reaches half the strength it has at the solenoid center. In deceleration mode, the solenoids are switched on before the synchronous molecule arrives in the solenoid, and switched off when the synchronous molecule has reached the position corresponding to $\phi_0$. In acceleration mode, the solenoid is switched on when the synchronous molecules has reached the position corresponding to $\phi_0$, and it is only switched off when the synchronous molecule no longer experiences the field induced by the solenoid. In hybrid mode, two adjacent solenoids are simultaneously activated to create a symmetric potential in the longitudinal direction. For this, each solenoid is activated twice: once when the synchronous molecule approaches, and once when the synchronous molecule exits the solenoid.

In this description we neglected the nonzero switching time of the current in the solenoids. In our decelerator, however, the current pulses feature a rise time of about 8~$\mu$s, as will be explained in more detail in section \ref{subsec:simulations}. In the simulations, the full current profile is taken into account; we will adopt the convention that the current has reached half of the maximum value when the synchronous particle reaches the $\phi_0$ position. This switching protocol ensures that in hybrid mode with $\phi_0=0^{\circ}$, the molecules will receive an equal amount of acceleration and deceleration, in analogy with operation of a Stark decelerator with $\phi_0=0^{\circ}$.

\begin{figure}[!htb]
    \centering
    \resizebox{1.0\linewidth}{!}
    {\includegraphics{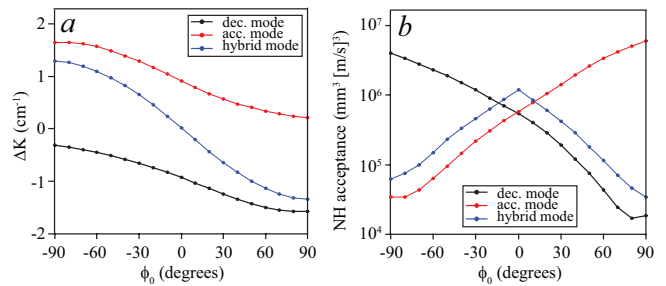}}
    \caption{(\emph{a}) Kinetic energy change $\Delta K$ per stage for a synchronous NH radical ($X\,^3\Sigma^-, N=0, J=1, M=1$), if the decelerator is operated in deceleration, acceleration and in hybrid mode. (\emph{b}) Corresponding 6D phase-space acceptances for NH radicals passing through a decelerator that consists of a suitable number of solenoids and hexapoles. }
    \label{fig:acceptance-overview}
\end{figure}

The kinetic energy change $\Delta K$ that the synchronous molecule experiences per stage is shown for each mode in Figure \ref{fig:acceptance-overview}\emph{a}. In this calculation we assume the NH radical in its electronic ground state, that has a 2-$\mu_B$ magnetic dipole moment (\emph{vide infra}). In the deceleration and acceleration modes, the full range of $\phi_0$ ($-90^{\circ}$ to $90^{\circ}$) can be used to reduce and increase the kinetic energy, respectively. In hybrid mode, deceleration and acceleration are achieved for $0^{\circ} < \phi_0 \leq 90^{\circ}$ and $-90^{\circ}\leq \phi_0 < 0^{\circ}$, respectively, whereas the packet is transported through the decelerator at constant speed for $\phi_0=0^{\circ}$. The maximum value for $\Delta K$ that can be achieved amounts to approximately 1.5~cm$^{-1}$.

\subsection{Numerical trajectory simulations}\label{subsec:simulations}
\begin{figure}[!htb]
    \centering
    \resizebox{1.0\linewidth}{!}
    {\includegraphics{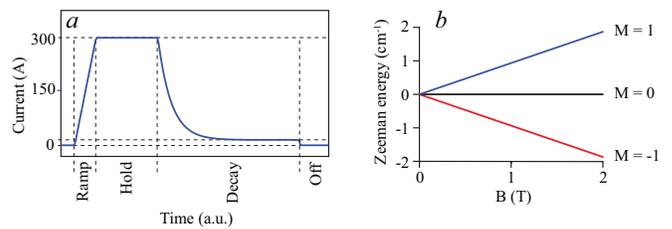}}
    \caption{(\emph{a}) Schematic representation of a single current pulse applied to a solenoid. (\emph{b}) Zeeman energy diagram for NH radicals in the $X\,^3\Sigma^-, N=0, J=1$ rotational ground state.}
    \label{fig:NHZeemanshift}
\end{figure}

The operation characteristics of the Zeeman decelerator are extensively tested using numerical trajectory simulations. In these simulations, it is essential to take the temporal profile of the current pulses into account. Unless stated otherwise, we assume single pulse profiles as illustrated in Figure \ref{fig:NHZeemanshift}\emph{a}. The current pulses feature a rise time of approximately 8~$\mu$s, then a variable hold-on time during which the current has a constant value of 300~A. The current exponentially decays to a lingering current of 15~A with a characteristic decay time of 5~$\mu$s, as can be created by switching the current to a simple resistor in the electronic drive unit. This lingering current is only switched off at much later times, and is introduced to prevent Majorana transitions as will be explained in section \ref{subsec:Majorana}. Furthermore, we assume that the hexapoles are always active when molecules are in their proximity.

In these simulations, we use NH radicals in the $X\,^3\Sigma^-, N=0, J=1$ rotational ground state throughout. The Zeeman effect of this state is shown in Figure \ref{fig:NHZeemanshift}\emph{b}. NH radicals in the low-field seeking $M=1$ component possess a magnetic moment of 2~$\mu_B$, and experience a linear Zeeman shift. NH radicals in this state have a relatively small mass-to-magnetic moment ratio of 7.5~amu/$\mu_B$, making NH a prime candidate for Zeeman deceleration experiments. Our findings are easily translated to other species by appropriate scaling of this ratio, in particular for species that also have a linear Zeeman shift (such as metastable helium, for instance).

\begin{figure}[!htb]
    \centering
    \resizebox{1.0\linewidth}{!}
    {\includegraphics{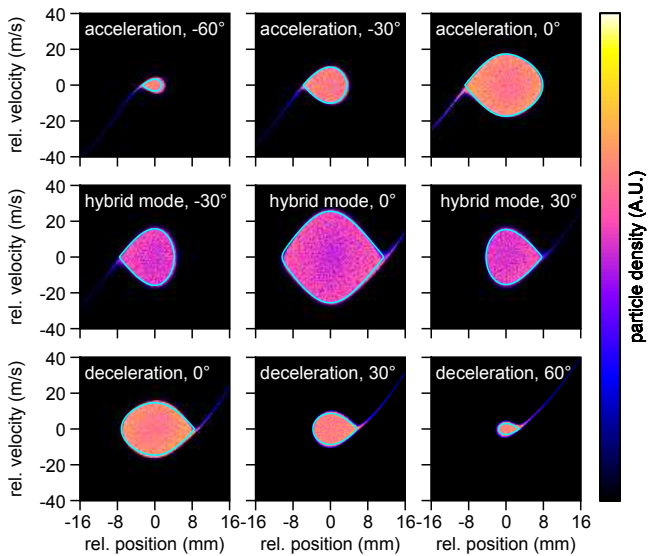}}
    \caption{\emph{Longitudinal} phase-space distributions of NH($X\,^3\Sigma^-, N=0, J=1$) in the final solenoid of the 100-stage decelerator when the decelerator is operated using  different switching modes and values for $\phi_0$. These distributions result from 3D trajectory simulations of 5$\cdot 10^{6}$ particles. Positions and velocities are given relative to the synchronous particle. The same normalization was used for all graphs. The theoretical separatrices for each sequence are visible as cyan overlays.}
    \label{fig:phasespace3D}
\end{figure}

The inherent 6D phase-space acceptance of the decelerator is investigated by uniformly filling a block-shaped area in 6D phase-space, and by propagating each molecule within this volume through a decelerator that consists of 100~solenoids and 100~hexapoles. In the range of negative $\phi_0$ in deceleration mode and positive $\phi_0$ for acceleration mode we instead used 200 pairs of solenoids and hexapoles to spatially separate the molecules within the phase stable area from the remainder of the distribution. This is explained in the appendix. The uniform distributions are produced using six unique Van der Corput sequences \cite{Corput:PAWA38:813}. For each of the three operation modes, the resulting longitudinal phase-space distributions of the molecules in the last solenoid of the decelerator are shown in Figure \ref{fig:phasespace3D} for three different $\phi_0$. The separatrices that follow from the 1D model for phase stability that explicitly takes the temporal profiles of the currents into account, as described in detail by Dulitz \emph{et al.} \cite{Dulitz:PRA91:013409}, are given as a cyan overlay.

In each simulation, the synchronous molecule has an initial velocity chosen such that the total flight time is approximately 4.8 ms. This results in velocity progressions of $[370 \rightarrow 625]$, $[390 \rightarrow 599]$ and $[421 \rightarrow 568]$ m/s in acceleration mode with $\phi_0=-60^{\circ}, -30^{\circ}$ and $0^{\circ}$, respectively; a progression of $[445 \rightarrow 550]$, $[500 \rightarrow 500]$ and $[550 \rightarrow 447]$ m/s in hybrid mode with $\phi_0=-30^{\circ}, 0^{\circ}$ and $30^{\circ}$; and finally a progression of $[570 \rightarrow 421]$, $[595 \rightarrow 399]$ and $[615 \rightarrow 383]$ m/s in deceleration mode corresponding to $\phi_0=0^{\circ}, 30^{\circ}$ and $60^{\circ}$.

It is shown that in all operation modes and for all values of $\phi_0$, the separatrices accurately describe the longitudinal acceptances of the decelerator. For larger values of $|\phi_0|$, the sizes of the separatrices are reduced, reflecting the smaller size and depth of the effective time-averaged potential wells. Note the symmetric shape of the separatrix when the decelerator is operated in hybrid mode with $\phi_0 = 0^{\circ}$, corresponding to guiding of the packet through the decelerator at constant speed. The transmitted particle density is slightly less in hybrid mode than in other modes, which indicates that the transverse acceptance is not completely independent of the solenoid fields. However, in each mode of operation the regions in phase-space accepted by the decelerator are homogeneously filled; no regions with a significantly reduced number of molecules are found. This is a strong indication that the decelerator indeed features phase stability.
\begin{figure}[!htb]
    \centering
    \resizebox{1\linewidth}{!}
    {\includegraphics{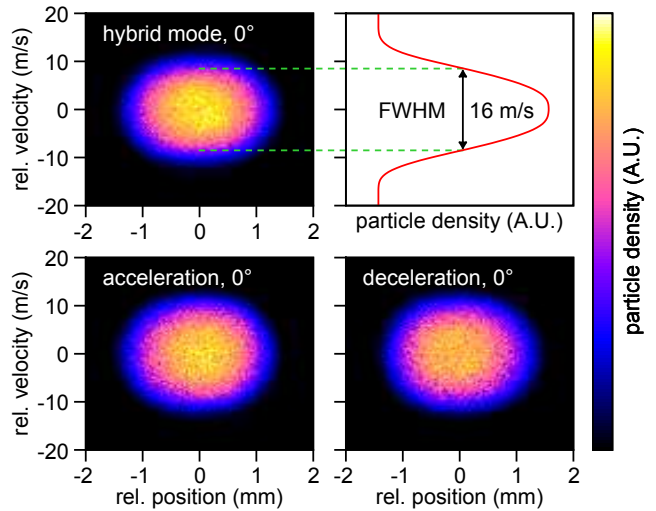}}
    \caption{\emph{Transverse} phase-space distributions that result from the same simulations used in Figure \ref{fig:phasespace3D}. The normalization is different for hybrid mode to compensate for the larger longitudinal acceptance. The top-right panel shows the transverse velocity profile for the hybrid mode, taken at the beam axis (0-mm transverse position).}
    \label{fig:transspace3D}
\end{figure}

The transverse acceptance is found to be rather independent of $\phi_0$, and is shown in Figure \ref{fig:transspace3D} for $\phi_0=0^{\circ}$ only. It can be seen that the transverse acceptance is typically smaller than the longitudinal acceptance, in accordance with our desideratum for molecular beam scattering experiments. Note that the transverse (velocity) acceptance can be modified independently from the deceleration and acceleration properties of the decelerator, simply by adjusting the field strength of the hexapoles.

Additionally, trajectory simulations can be used to quantify the overall 6D acceptance of the decelerator. Because of the uniform initial distribution, all particles that are propagated represent a small but equal volume in phase-space. At the end of the decelerator, the particles within a predefined range with respect to the synchronous particle are counted, yielding the volume in phase-space occupied by these particles. In the simulations, the initial "block" distribution is widened until the number of counted particles increases no further. We define the corresponding phase-space volume as the acceptance of the decelerator. The resulting 6D acceptance is shown for each operation mode in panel \emph{b} of Figure \ref{fig:acceptance-overview}.

Operating in hybrid mode results in the typical triangle-shaped acceptance curve as a function of $\phi_0$ that is also found for Stark decelerators. A maximum 6D phase-space acceptance of approximately $1.2 \cdot 10^6$ mm$^3$ (m/s)$^3$ is found for $\phi_0=0^{\circ}$, and drops below $10^5$ mm$^3$ (m/s)$^3$ at large $|\phi_0|$. A peculiar effect is seen in the deceleration and acceleration modes for $\phi_0<0^{\circ}$ and $\phi_0>0^{\circ}$, respectively. Here, the acceptance largely exceeds the acceptance for $\phi_0=0^{\circ}$, and approaches values of $6 \cdot 10^6$ mm$^3$ (m/s)$^3$. This is a special consequence of the continuously acting focusing forces of the hexapoles, and will be discussed in more detail in the Appendix.

Although one has to be careful to derive the merits of a decelerator from the 6D phase-space acceptance alone, it is instructive to compare these numbers to the phase-space acceptances found in other decelerators. Conceptually, the hybrid mode of our Zeeman decelerator is compared best to the $s=3$ mode of a Stark decelerator. For the latter, Scharfenberg \emph{et al.} found a maximum phase-space acceptance of $3 \cdot 10^5$ mm$^3$ (m/s)$^3$ for OH ($X\,^2\Pi_{3/2}, J=3/2$) radicals, with a similar partitioning of this acceptance between the longitudinal and transverse coordinates as found here \cite{Scharfenberg:PRA79:023410}. In comparison, for a multi-stage Zeeman decelerator without hexapoles, Wiederkehr \emph{et al.} found that the 6D acceptance peaks at about $2 \cdot 10^3$ mm$^3$ (m/s)$^3$ for Ne ($^3P_2$) atoms when equilibrium phase angles in the range $30^{\circ}$--$45^{\circ}$ degrees are used \cite{Wiederkehr:JCP135:214202}. The acceptance of the multi-stage Zeeman decelerator developed by Raizen and coworkers, also referred to as a magnetic coilgun, was reported to have an upper limit of $10^5$ mm$^3$ (m/s)$^3$ \cite{Narevicius:ChemRev112:4879}. The highest 6D acceptances to date are found in traveling wave decelerators, mostly thanks to the large transverse acceptances of these decelerators. The maximum acceptance of the traveling wave Zeeman decelerator of Narevicius and coworkers, for instance, amounts to $2 \cdot 10^7$ mm$^3$ (m/s)$^3$ for Ne ($^3P_2$) atoms \cite{Lavert-Ofir:PCCP13:18948}.

\subsection{Phase stability}
\begin{figure}[!htb]
    \centering
    \resizebox{0.5\linewidth}{!}
    {\includegraphics{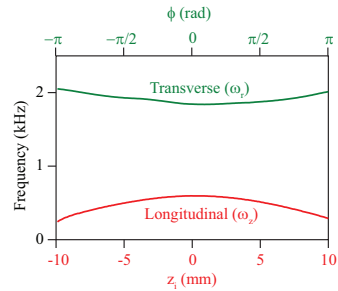}}
    \caption{Transverse and longitudinal oscillation frequencies of an NH ($X\,^3\Sigma^-, N=0, J=1$) radical as a function of the initial longitudinal position $z_i$ and phase $\phi$, respectively. The Zeeman decelerator is assumed to operate in hybrid mode with $\phi_0=0^{\circ}$. }
    \label{fig:frequencies}
\end{figure}

The numerical trajectory simulations yield very strong indications that the molecules are transported through the Zeeman decelerator without loss, i.e., phase stable operation is ensured. We support this conjecture further by considering the equation of motion for the transverse trajectories, using a model that was originally developed to investigate phase stability in Stark decelerators \cite{Meerakker:PRA73:023401}. In this model, we consider a (nonsynchronous) molecule with initial longitudinal position $z_i$ relative to the synchronous molecule, which oscillates in longitudinal phase-space around the synchronous molecule with longitudinal frequency $\omega_z$. In other words, during this motion the relative longitudinal coordinate $\phi$ oscillates around the synchronous value $\phi_0$. In the transverse direction, the molecule oscillates around the beam axis with transverse frequency $\omega_r$, which changes with $\phi$. In Figure \ref{fig:frequencies}, the longitudinal and transverse oscillation frequencies are shown that are found when the Zeeman decelerator is operated in hybrid mode with $\phi_0=0^{\circ}$. For deceleration and acceleration modes rather similar frequencies are found (data not shown). It can be seen that the transverse oscillation frequency largely exceeds the longitudinal oscillation frequency. As we will show below, this eliminates the instabilities that has deteriorated the phase-space acceptance of multi-stage Stark and Zeeman decelerators in the past \cite{Meerakker:PRA73:023401,Sawyer:EPJD48:197,Wiederkehr:JCP135:214202,Wiederkehr:PRA82:043428}.

\begin{figure}[!htb]
    \centering
    \resizebox{1.0\linewidth}{!}
    {\includegraphics{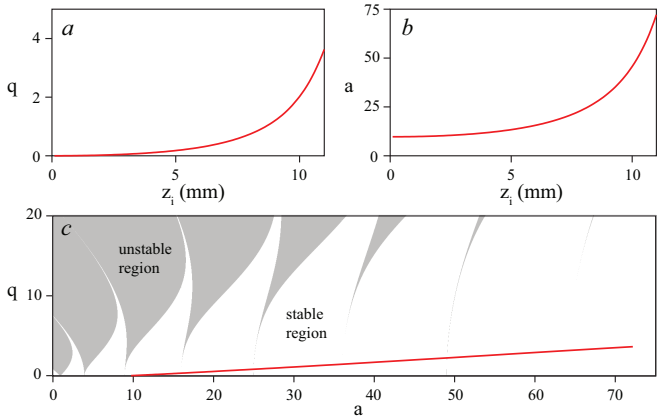}}
    \caption{Values for the parameters $q$ (panel \emph{a}) and $a$ (panel \emph{b}) for a nonsynchronous molecule as a function of the initial longitudinal position $z_i$ if the decelerator is operated in hybrid mode with $\phi_0=0^{\circ}$. The red curve in the Mathieu stability diagram (panel \emph{c}) characterizes the stability of the resulting trajectories. Based on the coupled equations of motion, trajectories are stable or unstable within the white or grey areas, respectively.}
    \label{fig:stability}
\end{figure}

During its motion, a molecule experiences a time-dependent transverse oscillation frequency that is given by \cite{Meerakker:PRA73:023401}:
\begin{equation}
\omega_r(t) = \omega^2_0-A \cos(2\omega_z t),
\label{eq:trans}
\end{equation}
where $\omega_0$ and $A$ are constants that characterize the oscillatory function. The resulting transverse equation of motion is given by the Mathieu differential equation:
\begin{equation}
\frac{d^2 r}{d\tau^2}+[a-2q\cos(2\tau)]r=0,
\end{equation}
with:
\begin{equation}
a=\left(\frac{\omega_0}{\omega_z}\right)^2, \qquad q=\frac{A}{2\omega^2_z}, \qquad \tau=\omega_z t.
\label{eq:param}
\end{equation}
Depending on the values of $a$ and $q$, the solution of this equation exhibits stable or unstable behavior. This is illustrated in Figure \ref{fig:stability} that displays the Mathieu stability diagram. Stable and unstable solutions exist for combinations of $a$ and $q$ within the white and gray areas, respectively. For each operation mode of the decelerator, and for a given phase angle $\phi_0$, the values for the parameters $a$ and $q$ can be determined from the longitudinal and transverse oscillation frequencies of Figure \ref{fig:frequencies}. The resulting values for the parameters $q$ and $a$ as a function of $z_i$ are shown in panel (\emph{a}) and (\emph{b}) of Figure \ref{fig:stability}, for the decelerator running in hybrid mode with $\phi_0=0^{\circ}$. The ($a,q$) combinations that govern the molecular trajectories for this operation mode are included as a solid red line in the stability diagram shown in panel (\emph{c}). Clearly, the red line circumvents all unstable regions, and only passes through the unavoidable "vertical tongues" where they have negligible width. These narrow strips do not cause unstable behavior for decelerators of realistic length. The unstable areas in the Mathieu diagram are avoided because of the high values of the parameter $a$. This same result was found for the other operation modes and equilibrium phase angles. We thus conclude that the insertion of hexapoles effectively decouples the transverse motion from the longitudinal motion; the Zeeman decelerator we propose is inherently phase stable, and can in principle be realized with arbitrary length.

\subsection{Prevention of Majorana losses}\label{subsec:Majorana}
An important requirement in devices that manipulate the motion of molecules using externally applied fields, is that the molecules remain in a given quantum state as they spend time in the device. As the field strength approaches zero, different quantum states may become (almost) energetically degenerate, resulting in a possibility for nonadiabatic transitions. This may lead to loss of particles, which is often referred to as Majorana losses.

The occurrence of nonadiabatic transitions has been studied extensively for neutral molecules in electric traps \cite{Kirste:PRA79:051401}, as well as for miniaturized Stark decelerators integrated on a chip \cite{Meek:PRA83:033413}. Tarbutt and coworkers developed a theoretical model based on the time-dependent Hamiltonian for the field-molecule interaction, and quantitatively investigated the transition probability as the field strength comes close to zero, and/or if the field vector rotates quickly relative to the decelerated particles \cite{Wall:PRA81:033414}.

In the multi-stage Zeeman decelerators that are currently operational, losses due to nonadiabatic transitions can play a significant role \cite{Hogan:PRA76:023412}. Specifically, when switching off a solenoid right as the particle bunch is near the solenoid center, there will be a moment in time where no well-defined magnetic quantization field is present. In previous multi-stage Zeeman decelerator designs, this was compensated by introducing a temporal overlap between the current pulses of adjacent solenoids, effectively eliminating nonadiabatic transitions \cite{Hogan:PRA76:023412}. In the Zeeman decelerator concept presented in this manuscript, this solution is not available, since adjacent solenoids are separated by hexapole elements. The hexapoles induce only marginal fringe fields, and do not contribute any magnetic field strength on the molecular beam axis.

\begin{figure}[!htb]
    \centering
    \resizebox{1.0\linewidth}{!}
    {\includegraphics{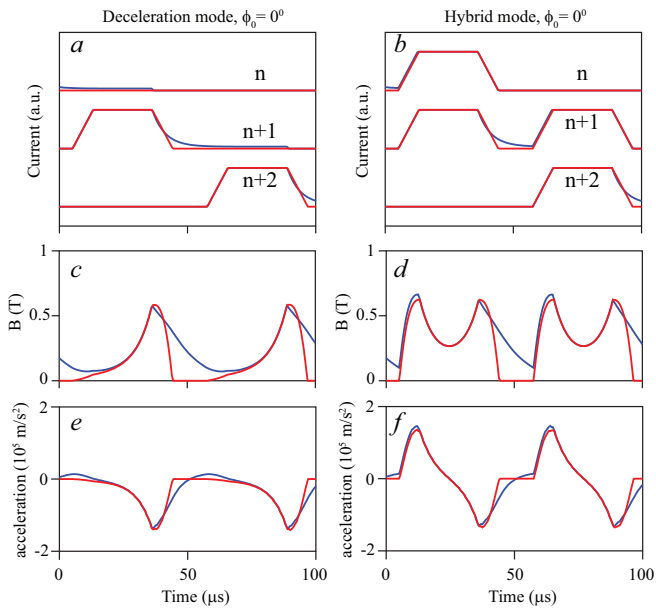}}
    \caption{Schematic representation of the switching protocol, when the Zeeman decelerator is operated in deceleration mode (left) and hybrid mode (right) with $\phi_0=0^{\circ}$. (\emph{a,b}) Sequence of current pulses applied to adjacent solenoids, using conventional ramp down profiles (red curves) or using a lingering decaying current (blue curves). (\emph{c,d}) Magnetic field strength experienced by the synchronous molecule as a function of time induced by the sequence of current pulses. (\emph{e,f}) Resulting acceleration of the synchronous molecule as a function of time.}
    \label{fig:Majorana-currents}
\end{figure}

Referring back to Figure \ref{fig:NHZeemanshift}\emph{a}, we introduce a quantization field throughout the hexapole-solenoid array by switching each solenoid to a low-level lingering current when the high current pulse is switched off. Since the fringe field of a solenoid extends beyond the geometric center of adjacent hexapoles, and since in the center of the solenoid the maximum magnetic field per unit of current is created, a lingering current of approximately 15~A is sufficient to provide a minimum quantization field of 0.1~T. The resulting sequences of current profiles through the solenoids with number $n$, $n+1$ and $n+2$ are shown in the upper half of Figure \ref{fig:Majorana-currents} for the deceleration (panel \emph{a}) and hybrid modes (panel \emph{b}). The profiles for acceleration mode are not shown here, but they feature the low current before switching to full current, instead of a low current after. The lingering current exponentially decays to its final value, and lasts until the next solenoid is switched off. In panels \emph{c} and \emph{d} the corresponding magnetic field strength is shown that is experienced by the synchronous molecule as it propagates through the decelerator (blue curves), together with the field that would have resulted if the solenoid were switched off with a conventional ramp time (red curves). Clearly, the low level current effectively eliminates the zero-field regions.

From model calculations similar to the ones developed by Tarbutt and coworkers \cite{Wall:PRA81:033414}, we expect that the magnetic field vector inside the solenoids will not rotate fast enough to induce nonadiabatic transitions, provided that all solenoid fields are oriented in the same direction. We therefore conclude that the probability for nonadiabatic transitions is expected to be negligible for the Zeeman decelerator concept proposed here.

One may wonder how the addition of the slowly decaying lingering current affects the ability to efficiently accelerate or decelerate the molecules. This is illustrated in panels (\emph{e}) and (\emph{f}) of Figure \ref{fig:Majorana-currents} that displays the acceleration rate experienced by the synchronous molecule. The acceleration follows from $-(\vec{\triangledown} U_{\textrm{Z}})/m$, where $U_{\textrm{Z}}$ is the Zeeman energy for NH ($X\,^3\Sigma^-, N=0, J=1, M=1$) induced by the time-varying magnetic field B(T), and $m$ is the mass of the NH radical. It is seen that the lingering current only marginally affects the acceleration force; a slight additional deceleration at early times is compensated by a small acceleration when the synchronous molecule exits the solenoid. Overall, the resulting values for $\Delta K$ with or without lingering current, as obtained by integrating the curves in panels (\emph{e}) and (\emph{f}), are almost identical (data not shown).

\subsection{Excessive focusing at low velocities}\label{subsec:low-velocities}
A common problem in multi-stage decelerators is the occurrence of losses due to excessive focusing at low forward velocities. This effect has been studied and observed in multi-stage Stark decelerators that operate in the $s=1$ or $s=3$ modes, where losses occur below approximately 50~or 150~m/s, respectively \cite{Sawyer:EPJD48:197,Scharfenberg:PRA79:023410}. Our concept for a multi-stage Zeeman decelerator shares these over-focusing effects at low final velocities, which may be considered a disadvantage over traveling wave decelerators, which are phase stable down to near-zero velocities.

At relatively high velocities, the hexapole focusing forces can be seen as a continuously acting averaged force, keeping the molecules confined to the beam axis. However, at low velocities this approximation is no longer valid, and the molecules can drift from the beam axis between adjacent hexapoles. We investigate the expected losses using similar numerical trajectory simulations as discussed in section \ref{subsec:simulations}, i.e., we again assume a Zeeman decelerator consisting of 100~hexapole-solenoid-pairs. We assume packets of molecules with five different mean initial velocities ranging between $v_{\textrm{in}}=$ 350~m/s and 550~m/s, and these packets are subsequently propagated through the decelerator. The decelerator is operated in hybrid mode, and can be used with different values for $\phi_0$. Since we assume a 100-stage decelerator throughout, the packets emerge from the decelerator with different final velocities.

\begin{figure}[!htb]
    \centering
    \resizebox{1.0\linewidth}{!}
    {\includegraphics{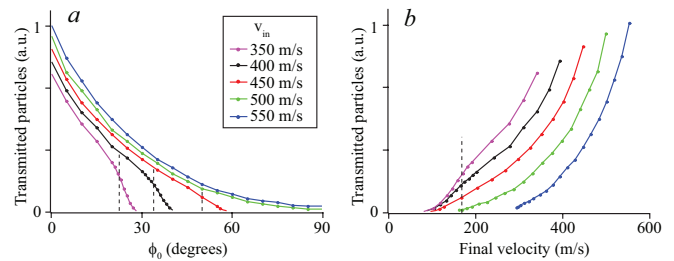}}
    \caption{Simulated number of transmitted NH radicals that pass through a 100~hexapole-solenoid-pair Zeeman decelerator operating in hybrid mode, as a function of \emph{$\phi_0$} (\emph{a}) and as a function of the final velocity (\emph{b}). Packets of NH are used with a mean initial velocity ranging between 350~m/s and 550~m/s.}
    \label{fig:over-focusing}
\end{figure}

In Figure \ref{fig:over-focusing} we show the number of decelerated particles that are expected at the end of the decelerator as a function of $\phi_0$ (panel \emph{a}), or as a function of the final velocity (panel \emph{b}). For low values of $\phi_0$, the transmitted number of molecules is (almost) equal for all curves; the slightly higher transmission for higher values of $v_{\textrm{in}}$ is related to the shorter flight time of the molecules in the decelerator. Consequently, molecules that are not within the inherent 6D phase-space acceptance of the decelerator can still make it to the end of the decelerator, and are counted in the simulations. For higher values of $\phi_0$, the transmitted number of molecules decreases, reflecting the reduction of the phase-space acceptance for these phase angles. This is particularly clear for the blue and green curves ($v_{\textrm{in}}=$ 550~and 500~m/s, respectively), which follow the 6D phase-space acceptance curve from Figure \ref{fig:acceptance-overview}\emph{b}. The three other curves feature a drop in transmission that occurs when the velocity drops below approximately 160~m/s, as is indicated by the dashed vertical lines. Obviously, for lower values of $v_{\textrm{in}}$, this velocity is reached at lower values of $\phi_0$ (see panel (\emph{a})).

The production of final velocities below this drop-off velocity is not a prime requirement in crossed beam scattering experiments, as the collision energy is determined by the velocities of both beams and the crossing angle between the beams. Very low collision energies can be reached using small crossing angles, relaxing the requirements on the final velocities of the reagent beams. For these applications we therefore see no direct need to combat these over-focusing effects. However, there are several promising options to mitigate these effects if desired. The first option is to employ hexapoles with a variable strength, such that the transverse oscillation frequency can be tuned along with the decreasing velocity of the molecular packet. Similarly, permanent hexapoles with different magnetization can be installed to modify the focusing properties. Finally, it appears possible to merge a hexapole and solenoid into a single element, by superimposing a hexapole arrangement on the outer diameter of the solenoid. Although technically more challenging, this approach will provide an almost continuously acting transverse focusing force, while keeping the possibility to apply current pulses to the solenoids. Preliminary trajectory simulations suggest that indeed a significant improvement can be achieved, but the validity of these approaches will need to be investigated further if near-zero final velocities are required.

\section{Experimental implementation}\label{sec:experiment}

\subsection{Multi-stage Zeeman decelerator}
\begin{figure}[!htb]
    \centering
    \resizebox{1.0\linewidth}{!}
    {\includegraphics{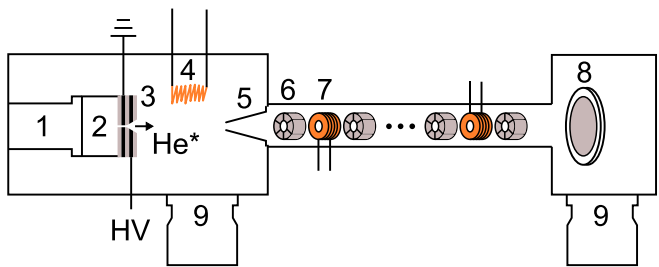}}
    \caption{Schematic representation of the experimental setup. The main components are: 1, coldhead; 2, Even-Lavie valve; 3, pinhole discharge; 4, hot filament; 5, skimmer; 6, hexapole; 7, solenoid; 8, MCP detector; 9, turbo-molecular pump}
    \label{fig:schematic_setup}
\end{figure}

An overview of the experiment is shown in Figure \ref{fig:schematic_setup}. The generation and detection of the metastable helium beam will be discussed in section \ref{subsec:He}; in this section we will first describe the decelerator itself, starting with a description of the solenoids and associated electronics.

\begin{figure*}[!htb]
    \centering
    \resizebox{1.0\linewidth}{!}
    {\includegraphics{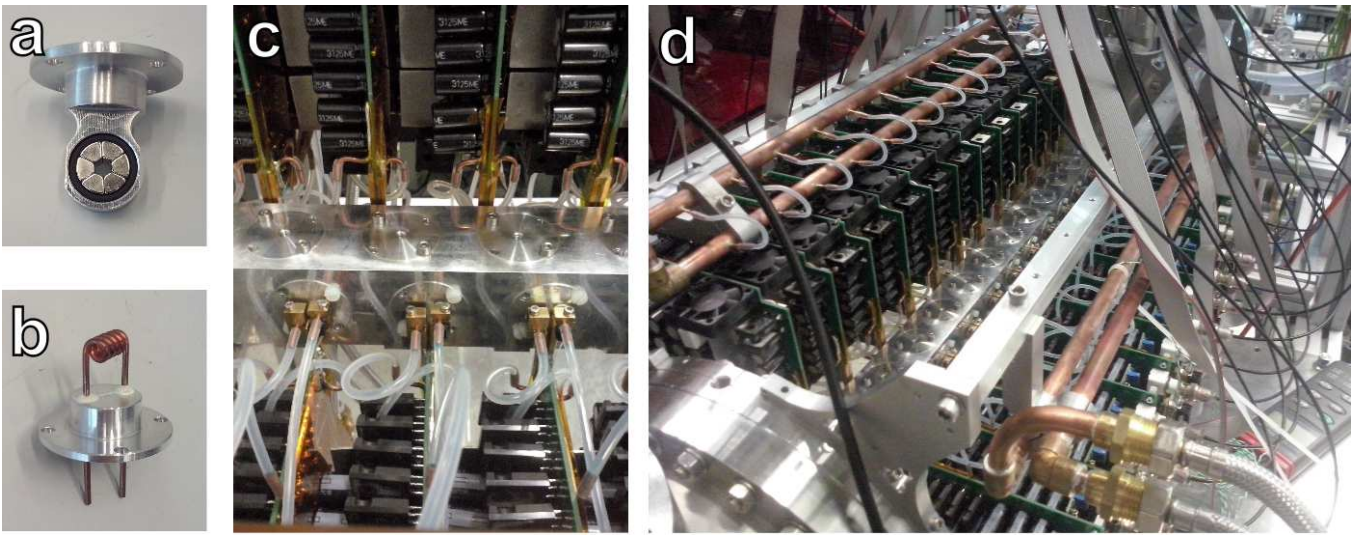}}
    \caption{Photographs of the Zeeman decelerator and the most relevant components. (\emph{a}) A hexapole element embedded in a custom flange; (\emph{b}) a solenoid element with connections through a custom flange; (\emph{c}) close-up of the decelerator housing; (\emph{d}) an overview of the decelerator. }
    \label{fig:setup_photo}
\end{figure*}

An essential part in a multi-stage Zeeman decelerator is the design of the deceleration solenoids, and the cooling strategy to remove the dissipated energy. A variety of solenoid designs have been implemented successfully in multi-stage Zeeman decelerators before. Merkt and coworkers utilized tightly-wound solenoids of insulated copper wire that were thermally connected to water-cooled ceramics \cite{Vanhaecke:PRA75:031402}. Later, similar solenoids were placed outside a vacuum tube, and submerged in a bath of cooling water \cite{Hogan:JPB41:081005}. This improved the cooling capacity, and enabled the experiment to operate at repetition rates of 10 Hz. Raizen and coworkers also developed a multi-stage Zeeman decelerator, referred to as the atomic or molecular coilgun, that is based on solenoids encased in high permeability material to increase the on-axis maximum magnetic field strength \cite{Narevicius:PRA77:051401,Liu:PRA91:021403}. Recently, different types of traveling wave Zeeman decelerators have been developed, which consist of numerous spatially overlapping quadrupole solenoids \cite{Lavert-Ofir:PCCP13:18948}, or a helical wire arrangement to produce the desired magnetic field \cite{Trimeche:EPJD65:263}.

In the decelerator presented here, we use a new type of solenoid that is placed inside vacuum, but that allows for direct contact of the solenoid material with cooling liquid. The solenoids consist of 4 windings of a copper capillary that is wound around a 3~mm bore diameter. The capillary has an inner diameter of 0.6~mm and an outer diameter of 1.5~mm, and cooling liquid is circulated directly through the capillary. The solenoid is wound such that the first and last windings end with a straight section of the capillary, as is shown in a photograph of a single solenoid in Figure \ref{fig:setup_photo}\emph{b}. These straight sections are glued into an aluminum mounting flange, as will be further discussed later. The inherent magnetic field profile generated by this solenoid is very similar to the solenoids as used in the simulations presented in section \ref{sec:concept}.

The use of a single layer of rather thick copper capillary as solenoid material in a Zeeman decelerator is unconventional, but it has some definite advantages. Because of the low-resistance copper capillary, small operating voltages (24 V) are sufficient to generate currents of approximately 4.5 kA that produce a maximum field of 2.2 T on the solenoid axis. This in turn allows for the use of FET-based electronics components to switch these currents, which are considerably cheaper than their high voltage IGBT-based counterparts. The same holds for the power supplies that deliver the current. By running cooling liquid directly through the solenoid capillary, the solenoids are efficiently cooled. The low operation voltage ensures that the cooling liquid does not conduct any significant electricity.

\begin{figure*}[!htb]
    \centering
    \resizebox{0.8\linewidth}{!}
    {\includegraphics{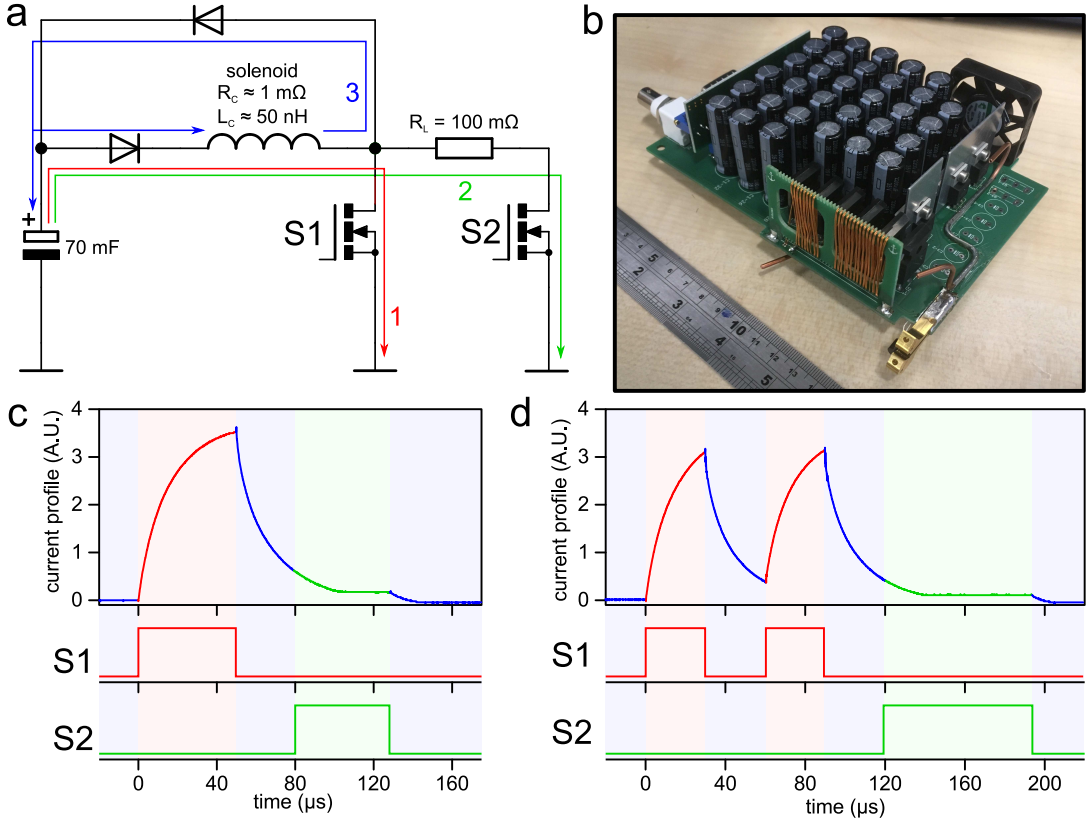}}
    \caption{Schematic representation of the driving electronics and the current pulse through a solenoid. (\emph{a}) A schematic drawing of the electronic circuit enabling the solenoid current. The colored pathways correspond to: 1 (red) maximum current through solenoid, 2 (green) reduced current through solenoid, and 3 (blue) dissipation of any leftover current in the solenoid. (\emph{b}) Photograph of a circuit board driving an individual coil. Current profile of a single (\emph{c}) or double (\emph{d}) pulse through the solenoid derived from the voltage induced in a smaller pickup solenoid. The color coding in the profile indicates the electronic pathway in (a). The trigger pulses for gates S1 and S2 to generate these pulses are shown underneath the current profiles. }
    \label{fig:electronics}
\end{figure*}

The current pulses are provided by specially designed circuit boards; one such board is displayed in Figure \ref{fig:electronics}\emph{b}. Each solenoid is connected to a single board, that is mounted directly onto the solenoid-flange feedthroughs in order to minimize power loss between board and solenoid. Brass strips are used to mechanically clamp the board to the capillary material. The simplified electronic circuit is shown schematically in Figure \ref{fig:electronics}\emph{a}. The circuit board is mostly occupied by a parallel array of capacitors, with a total capacitance of 70~mF. The capacitors are charged by a 24~V power supply and then discharged through the connected solenoid. The solenoids have a very low resistance $R_C$ of about 1~m$\Omega$ and self-inductance $L_C$ of about 50~nH, even compared to the electronic circuit itself. The capacitors are discharged via two possible pathways indicated in red and green, respectively, by activating the two independent gates S1 and S2. Closing gate S1 will allow electrons to flow through the solenoid, generating a maximum current of about 4.5 kA. Closing gate S2 will send the flow through both the solenoid and a 100 m$\Omega$ resistor that limits the current to about 150 A. When both gates are opened any remaining power in the solenoid will either dissipate in the electrical components along pathway 3 (in blue) or return to the capacitors. The electronic configuration is able to apply up to two consecutive pulses to each solenoid, as is required for the hybrid mode of operation.

As an example, the current profiles for a single pulse or double pulse are shown in Figure \ref{fig:electronics}\emph{c} and \ref{fig:electronics}\emph{d}, respectively, together with the trigger pulses that activate gates S1 and S2. These profiles were obtained from the induced voltage over a miniature solenoid that was placed inside the center of a decelerating solenoid \cite{Wiederkehr:JCP135:214202}. The current pulse is initiated by closing gate S1, after which the solenoid current shows a rapid rise to a maximal current of approximately 4.5~kA. After reopening gate S1, the current exponentially decreases with a time constant of 10~$\mu$s. Gate S2 is programmed to close automatically for a fixed duration of 50~$\mu$s, starting 30 $\mu$s after the reopening of S1. While gate S2 is closed, a low-level lingering current is maintained in the solenoid to prevent Majorana transitions (see section \ref{subsec:Majorana}), providing a quantization field for atoms or molecules that are near the solenoid.

The solenoids and electronics boards are actively cooled using a closed-cycle cooling system. An approximately 10-cm-long capillary section is soldered onto each electronics board, and each capillary is connected in series to its connecting solenoid using silicon tubes. All board-solenoid pairs are individually connected to a mains and return cooling line, using the same flexible silicon material. Each electronics board is additionally cooled by a small fan. Using this cooling system, relatively low operation temperatures are maintained despite the high currents that are passed through the solenoid. In the experiments shown here, the Zeeman decelerator is routinely operated with a repetition frequency of 10~Hz, while the temperature of the solenoids is kept below 40~degrees Celsius.

The solenoids are pulsed in a predefined time sequence designed to control the longitudinal velocity of a specific paramagnetic particle. This time sequence is calculated while taking current profiles into account that are modeled after the measured profiles shown in Figure \ref{fig:electronics}\emph{b}. The resulting pulse sequence for gates S1 and S2 is programmed into a pattern generator (Spincore PulseBlaster PB24-100-4k), which sends pulse signals to each individual circuit board. The temperature of each solenoid is continuously monitored via a thermocouple on the connecting clamps of the circuit board. When the temperature of the solenoid exceeds a user-set threshold, operation of the decelerator is interrupted.

The magnetic hexapoles consist of six wedge-shaped permanent magnets in a ring, as seen in Figure \ref{fig:setup_photo}\emph{a}. Adjacent magnets in the ring have opposite radial remanence. The inner diameter of the hexapole is 3~mm and the length is 8~mm, such that these dimensions match approximately to the corresponding solenoid dimensions. The magnets used in this experiment are based on NdFeB (grade N42SH) with a remanence of approximately 450~mT. The advantage of using hexapoles consisting of permanent magnets is twofold: first, implementation is mechanically straightforward, and second, no additional electronics are needed to generate the focusing fields. However, this approach lacks any tunability of the field strength. This can in part be overcome by selectively removing hexapoles from the decelerator, or by exchanging the magnets for ones with a different magnetization. If required, electromagnetic hexapoles that allow for tunability of the field strength can be used instead. We have built and successfully operated hexapoles that are made of the solenoid capillary material, and could optimize their focusing strength by simply adjusting the time these hexapoles are switched on. However, we found that similar beam densities were achieved using the permanent hexapoles, and experiments with electromagnetic hexapoles are not further discussed here.

The decelerator contains 24 solenoids and 25 hexapoles that are placed with a center-to-center distance of 11~mm inside a vacuum chamber. The chamber consist of a hollow aluminum block of length 600~mm with a squared cross section of side lengths 40~mm. This chamber is made by machining the sides of standard aluminum pipe material with an inner diameter of 20~mm. Solenoids and hexapoles are mounted on separate flanges, as can be seen in Figure \ref{fig:setup_photo}, such that each element can be installed or removed separately. The first and last element of the decelerator is a hexapole to provide transverse focusing forces at the entrance and exit of the decelerator, respectively. Openings for the individual flanges on the decelerator housing spiral along the sides between subsequent elements, with clockwise 90~degree rotations. In this way there is enough space on each side of the decelerator to accommodate the electronics boards of the solenoids, which have a 42~mm height. In addition, since subsequent solenoids are rotated by 180~degrees in the decelerator, any asymmetry in the magnetic field because of the relatively coarse winding geometry is compensated.

Vacuum inside the decelerator housing is maintained by a vacuum pump installed under the detection chamber, which has an open connection to the decelerator housing. Only a minor pressure increase in the chamber is observed if the solenoids are operational, reflecting the relatively low operational temperature of the solenoids. Although for long decelerators additional pumping capacity inside the decelerator is advantageous, we find that for the relatively short decelerator used here the beam density is hardly deteriorated by collisions with background gas provided the repetition rate of the experiment is below 5~Hz. Under these conditions, the pressure in the decelerator maintains below $5 \cdot 10^{-7}$ mbar.

\subsection{Metastable helium beam}\label{subsec:He}
A beam of helium in the metastable (1s)(2s) $^3S$ ($m_S = 1$) state (from this point He*) was used to test the performance of the Zeeman decelerator. This species was chosen for two main reasons. First, He* has a small mass-to-magnetic-moment ratio (2.0 amu/$\mu_B$) with a large Zeeman shift, which allows for effective manipulation of the atom with magnetic fields. This allows us to significantly vary the mean velocity of the beam despite the relatively low number of solenoids. Second, He* can be measured directly with a micro-channel plate (MCP) detector, without the need for an ionizing laser, such that full time-of-flight (TOF) profiles can be recorded in a single shot. This allows for a real-time view of TOF profiles when settings of the decelerator are changed, and greatly facilitates optimization procedures.

The beam of He* is generated by expanding a pulse of neat He atoms into vacuum using a modified Even-Lavie valve (ELV) \cite{Even:JCP112:8068} that is cooled to about 16~K using a commercially available cold-head (Oerlikon Leybold). At this temperature the mean thermal velocity of helium is about 460~m/s. The ELV nozzle is replaced by a discharge source consisting of alternating isolated and conducting plates, similar to the source described by Ploenes \emph{et al.} \cite{Ploenes:RSI87:053305}. The discharge occurs between the conducting plates, where the front plate is kept at -600~V and the back plate is grounded. To ignite the discharge, a hot filament running 3~A of current is used. The voltage applied to the front plate is pulsed (20-30 $\mu$s duration) to reduce the total energy dissipation in the discharge. Under optimal conditions, a beam of He* is formed, with a mean velocity just above 500~m/s. Unless stated otherwise, in the experiments presented here, the decelerator is programmed to select a packet of He* with an initial velocity of 520~m/s.

The beam of He* passes through a 3~mm diameter skimmer (Beam Dynamics, model 50.8) into the decelerator housing. The first element (a hexapole) is positioned about 70~mm behind the skimmer orifice. The beam is detected by an MCP detector that is positioned 128~mm downstream from the exit of the decelerator. This MCP is used to directly record the integrated signal from the impinging He* atoms.

\section{Results and Discussion} \label{sec:results}
\subsection{Longitudinal velocity control}

As explained in section \ref{sec:concept}, the decelerator can be operated in three distinct modes of operation: in deceleration or acceleration modes, the atoms are most efficiently decelerated or accelerated, respectively, whereas in the so-called hybrid mode of operation, the beam can be transported or guided through the decelerator at constant speed (some mild deceleration or acceleration is in principle also possible in this mode). In this section, we present experimental results for all three modes of operation.

\begin{figure}[!htb]
    \centering
    \resizebox{1.0\linewidth}{!}
    {\includegraphics{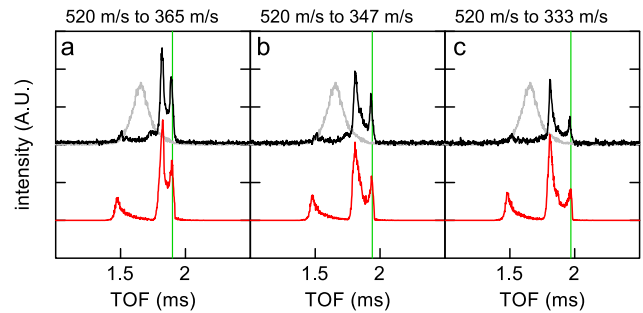}}
    \caption{TOF profiles of He* exiting the Zeeman decelerator when the decelerator is operated in deceleration mode using an effective equilibrium phase angle of 30$^{\circ}$ (\emph{a}), 45$^{\circ}$ (\emph{b}) and 60$^{\circ}$ (\emph{c}). Deceleration sequences were used to select an initial velocity of 520 m/s, resulting in the final velocities mentioned above each graph. Black traces are the experimentally obtained profiles, red traces show the profiles that result from numerical trajectory simulations. The gray traces show the TOF profiles that are measured when all solenoids are switched off. Vertical green lines indicate the arrival times of the synchronous atoms that are expected from the simulations. }
    \label{fig:decelTOF}
\end{figure}

We will start with the regular deceleration mode. In Figure \ref{fig:decelTOF}, TOF profiles for He* atoms exiting the decelerator are shown that are obtained when the decelerator is operated in deceleration mode, using different values for the equilibrium phase angle $\phi_0$. In the corresponding pulse sequences, the synchronous atom is decelerated from 520~m/s to 365~m/s, 347~m/s and~333 m/s, corresponding to effective equilibrium phase angles of 30$^{\circ}$, 45$^{\circ}$ and 60$^{\circ}$, respectively. The corresponding loss of kinetic energy amounts to 23~cm$^{-1}$, 25~cm$^{-1}$ and 27~cm$^{-1}$. The arrival time of the synchronous atom in the graphs is indicated by the vertical green lines. Black traces show the measured profiles; the gray traces that are shown as an overlay are obtained when the decelerator was not operated, i.e., the solenoids are all inactive but the permanent hexapole magnets are still present to focus the beam transversely.

The experimental TOF profiles are compared with profiles that result from three dimensional trajectory simulations. In these simulations, an initial beam distribution is assumed that closely resembles the He* pulse generated by the modified ELV. The resulting TOF profiles are shown in red, vertically offset from the measured profiles for clarity. The simulated profiles show good agreement with the experiment, both in relative intensity and arrival time of the peaks. However, it must be noted that the relative intensities are very sensitive to the chosen parameters of the initial He* pulse. By virtue of the supersonic expansion and discharge processes, these distributions are often not precisely known, and may vary from day to day. Nevertheless, the agreement obtained here, in particular regarding the overall shape of the TOF profiles and the predicted arrival times of the decelerated beam, suggests that the trajectory simulations accurately describe the motion of atoms inside the decelerator. No indications are found for unexpected loss of atoms during the deceleration process, or for behavior that is not described by the simulations.

\begin{figure}[!htb]
    \centering
    \resizebox{1.0\linewidth}{!}
    {\includegraphics{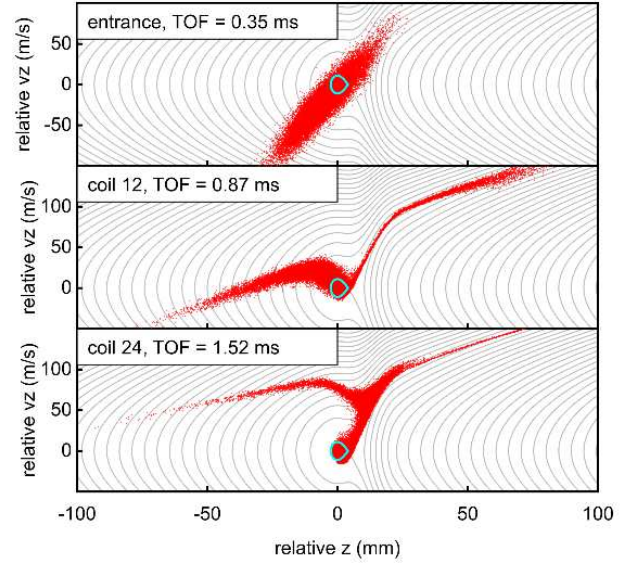}}
    \caption{Longitudinal phase-space distributions of He* atoms during a deceleration sequence from 520 m/s to 365 m/s, obtained from simulations. The horizontal (vertical) axis shows the longitudinal position (velocity) relative to the synchronous atom. Each red dot represents a simulated He* atom. Grey contour lines show the expected trajectories derived from the averaged Zeeman potential energy, and the cyan contour shows the separatrix of the stable region in phase-space. The distributions are sampled at the time the synchronous atom enters the decelerator (top), when it arrived at solenoid 12 (middle) and solenoid 24 (bottom).}
    \label{fig:decelphase}
\end{figure}

The profiles presented in Figure \ref{fig:decelTOF} show more features than the decelerated packets alone. In particular, there is an additional decelerated peak in each of the graphs that is more intense but slightly faster than the decelerated packet. We use the three dimensional trajectory simulations to study the origin of this feature. In Figure \ref{fig:decelphase}, the longitudinal phase-space distributions are shown that result from these simulations at the entrance (upper panel), middle (central panel), and exit (lower panel) of the decelerator. The simulation pertains to the situation that results in the TOF profiles presented in Figure \ref{fig:decelTOF}\emph{a}, i.e., the decelerator is operated in deceleration mode with $\phi_0=30^{\circ}$.

In these phase-space distributions, the grey contour lines depict the predicted trajectories considering the time-averaged Zeeman potential energy. The separatrix of the stable phase-space is highlighted with a cyan overlay. From this evolution of the longitudinal phase-space distribution, we can understand the origin of various pronounced features in the TOF profiles. The first peak in each TOF profile is a collection of the fastest particles in the initial beam distribution. These particles are hardly affected by the solenoids, and propagate to the detector almost in free flight. However, the part of the beam that is initially slower than the synchronous molecule is strongly affected by the solenoids. This part eventually gains in velocity relative to the decelerated bunch, resulting in an ensemble of particles with a relatively high density. This part arrives at the detector just before the decelerated He* atoms, resulting in the second intense peak in the TOF profiles of Figure \ref{fig:decelTOF}. It is noted that this peak appears intense because our decelerator is rather short, leaving insufficient time for the decelerated bunch to fully separate from the initial beam distribution. For longer decelerators, the part of the beam that is not enclosed by the separatrix will gradually spread out, and its signature in the TOF profiles will weaken.

The phase-space distributions that are found at the end of the decelerator may also be used to determine the velocity width of the decelerated packet of atoms. For the examples of Figure \ref{fig:decelphase}, these widths are about 25~m/s.

\begin{figure}[!htb]
    \centering
    \resizebox{1.0\linewidth}{!}
    {\includegraphics{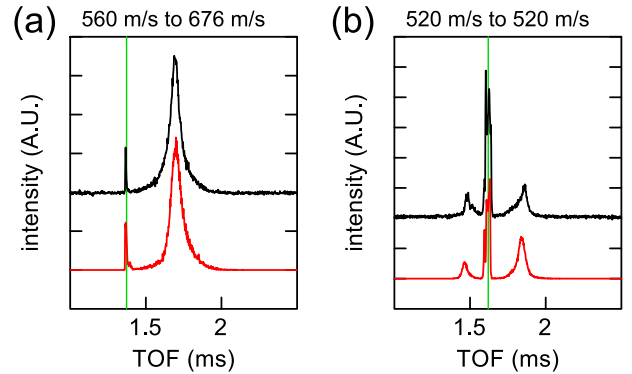}}
    \caption{(\emph{a}) TOF profile of He* accelerated by a multi-stage Zeeman decelerator from 560 m/s to 676 m/s. The black and red trace correspond to the experimental and simulated TOF profiles, respectively. (\emph{b}) Experimental (black) and simulated (red) TOF profile when the decelerator operates in hybrid mode, guiding a bunch of atoms with mean velocity of 520~m/s. The vertical green lines depict the arrival time of the synchronous atom in the simulation.}
    \label{fig:accelTOF}
\end{figure}

For completeness, we also measure a TOF profile when the decelerator is operated in acceleration mode. Figure \ref{fig:accelTOF}\emph{a} shows the TOF profile for the acceleration of He* atoms from an initial velocity of 560~m/s to a final velocity of 676 m/s. The simulated profile (red trace) shows good agreement with the experimental profile (black trace). Again, the vertical green line indicates the expected arrival time of the accelerated bunch. The sequence selects the fastest atoms in the beam, which is why no additional peaks are visible.

Finally, we study the performance of the decelerator in hybrid mode. This mode of operation allows for guiding of the beam through the decelerator at constant speed. In Figure \ref{fig:accelTOF}\emph{b} a TOF profile is shown when the decelerator is operated in hybrid mode and $\phi_0=0^{\circ}$, selecting an initial velocity of 520~m/s. The simulated TOF profile (red trace) again shows good agreement with the experimental TOF profile (black trace), although the intensity ratios between the guided part and the wings of the distributions are slightly different in the simulations than in the experiment. This is attributed to the idealized initial atom distribution that are assumed in the simulations.

\subsection{Presence of metastable helium molecules}

While our experiment is designed to decelerate He* atoms in the $^3S$ state, other types of particles may be created in the discharged beam as well. Specifically, formation of metastable He$_2$ molecules in the a$^3\Sigma$ state (from here on He$_2$*) is expected, as is also observed in the experiments by Motsch \emph{et al.} and Jansen \emph{et al.} that use a similar discharge source \cite{Motsch:PRA89:043420,Jansen:PRL115:133202}. However, He$_2$* is indistinguishable from He* in our detection system. In order to probe both species separately, mass selective detection using a non-resonant laser ionization detection scheme is used. Ultraviolet (UV) laser radiation with a wavelength of 243~nm is produced by doubling the light from an Nd:YAG-pumped pulsed dye laser running with Coumarin 480~dye, and focused into the molecular beam close to the exit of the decelerator. The resulting ions are extracted with an electric field of about 1 kV/cm and accelerated towards an MCP detector, where the arrival time of the ions reflect their mass over charge ratio.

\begin{figure}[!htb]
    \centering
    \resizebox{0.8\linewidth}{!}
    {\includegraphics{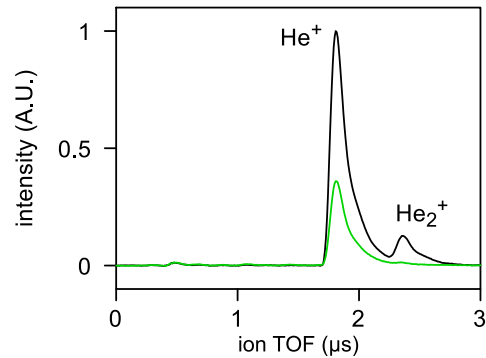}}
    \caption{Ion-TOF spectra of ions created when UV radiation interacts with the He* beam that exit the Zeeman decelerator at various times. The black trace is taken for a beam that passes through the decelerator with inactive coils, while the green trace results from a Zeeman decelerated part of the beam.}
    \label{fig:Hespec}
\end{figure}

We used this detection scheme to investigate the chemical composition of the beam that exits the Zeeman decelerator. In Figure \ref{fig:Hespec}, ion TOF spectra (i.e., the arrival times of the ions at the MCP detector with respect to the laser pulse) are shown. The black trace shows the ion TOF spectrum if the beam of He* atoms is passed through the decelerator without operating the solenoids. The UV laser is fired at the mean arrival time of the beam in the laser ionization region. Two peaks are clearly visible corresponding to the expected arrival time of He$^+$ and He$_2^+$, confirming that indeed He$_2$ molecules are created in the discharge. He atoms and molecules are detected in an 8:1 ratio in the neutral beam.

The green trace in Figure \ref{fig:Hespec} shows the ion TOF spectra that is recorded when the solenoids are operated for a typical deceleration sequence similar to the ones used to generate Figure \ref{fig:decelTOF}. This trace was taken when the UV laser selectively detects the decelerated part of the He* beam. Here, only He$^+$ is present in the ion TOF spectrum. Although He$_2$* has the same magnetic moment as He* and will thus experience the same force, the double mass of the molecule results in only half the acceleration. He$_2$* is therefore not decelerated at the same rate as He*, and will not exit the decelerator at the same time as the decelerated He* atoms. In conclusion, the Zeeman decelerator is quite effective in separating He* from the He$_2$*; the decelerated bunch only contains those species and/or particles in the quantum level for which the deceleration sequence was calculated.

\begin{figure}[!htb]
    \centering
    \resizebox{0.8\linewidth}{!}
    {\includegraphics{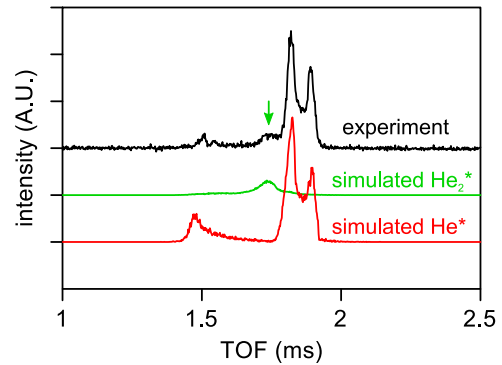}}
    \caption{TOF profile from Figure \ref{fig:decelTOF}\emph{a} revisited. The experimental (black) profile is compared to simulated profiles for He* (red) and He$_2$* (green). The simulated signal of He$_2$* is scaled to 1/8 of the He* signal. Each profile is vertically offset for reasons of clarity.}
    \label{fig:Hematch}
\end{figure}

Referring back to Figures \ref{fig:decelTOF} and \ref{fig:accelTOF} that were recorded without laser-based mass spectroscopic detection, one may wonder how the presence of He$_2$ molecules in the beam affect the recorded TOF profiles. Figure \ref{fig:Hematch} revisits the measurement from Figure \ref{fig:decelTOF}\emph{a}, but taking also He$_2$ molecules into account with the appropriate ratio to generate the simulated TOF profile. The resulting TOF profile for He$_2$* molecules is shown by the green trace, and is seen to fill the part of the TOF that was under represented by the original simulations (indicated by the vertical green arrow for the experimental trace).

\section{Conclusions and Outlook}
We have presented a new type of multi-stage Zeeman decelerator that is specifically optimized for scattering experiments. The decelerator consists of an array of alternating solenoids and hexapoles, that effectively decouples the longitudinal deceleration and transverse focusing forces. This ensures that phase-stable operation of the decelerator is possible over a wide range of velocities. For applications in scattering experiments, this decelerator concept has a number of advantages over existing and experimentally demonstrated Zeeman decelerators. The decelerator can be operated in three distinct modes that make either acceleration, deceleration, or guiding at constant speed possible, enabling the production of molecular packets with a continuously tunable velocity over a wide range of final velocities. Phase stability ensures that molecules can be transported through the decelerator with minimal loss, resulting in relatively high overall 6D phase-space acceptance. Most importantly, this acceptance is distributed unequally between the longitudinal and transverse directions. Both the spatial and velocity acceptances are much larger in the longitudinal than in the transverse directions, which meets the requirements for beam distributions in scattering experiments in an optimal way. At low final velocities, however, losses due to over-focusing occur. In crossed beam scattering experiments this appears inconsequential, but for trapping experiments---where low final velocities are essential---the use of the concept presented here should be carefully considered. We have discussed various promising options for combating these losses using alternative hexapole designs in the last section of the decelerator. Additionally, Zhang \emph{et al.} recently proposed a new operation scheme in a Stark decelerator that optimizes the transmitted particle numbers and velocity distributions, which could potentially be translated to a Zeeman decelerator \cite{Zhang:PRA93:023408}. The validity of these approaches will need to be investigated further, especially if near-zero final velocities are required.

In a proof-of-principle experiment, we demonstrated the successful experimental implementation of a new concept presented here, using a decelerator that consist of 24 solenoids and 25 hexapoles. The performance of the decelerator was experimentally tested using beams of metastable helium atoms. Both deceleration, acceleration, and guiding of a beam at constant speed have been demonstrated. The experimental TOF profiles of the atoms exiting the decelerator show excellent agreement with the profiles that result from numerical trajectory simulations. Although the decelerator presented here is relatively short, up to 60\% of the kinetic energy of He* atoms that travel with an initial velocity of about 520~m/s could be removed.

In the Zeeman decelerator presented here, we utilize a rather unconventional solenoid design that uses a thick copper capillary through which cooling liquid is circulated. The solenoid design allows for the switching of high currents up to 4.5~kA, using readily available and cheap low-voltage electronics components. The design is mechanically simple, and can be built at relatively low cost. We are currently developing an improved version of the decelerator, that is fully modular, and which can be extended to arbitrary length. The modules can be connected to each other without mechanically disrupting the solenoid-hexapole sequence, while the housing design will allow for the installation of sufficient pumping capacity to maintain excellent vacuum conditions throughout the decelerator. Operation of the Zeeman decelerator consisting of 100 solenoids and 100 hexapoles at repetition rates up to 30~Hz appears technically feasible.

\section{Acknowledgments}
The research leading to these results has received funding from the European Research Council under the European Union's Seventh Framework Programme (FP7/2007-2013/ERC grant agreement nr. 335646 MOLBIL). This work is part of the research program of the Netherlands
Organization for Scientific Research (NWO). We thank Katrin Dulitz, Paul Janssen, Hansj{\"u}rg Schmutz and Fr{\'e}d{\'e}ric Merkt for stimulating discussions on Zeeman deceleration, solenoid focusing properties and current switching protocols. We thank Rick Bethlem and Fr{\'e}d{\'e}ric Merkt for carefully reading the manuscript and for valuable suggestions for textual improvements. We thank Gerben Wulterkens for the design of prototypes.

\section{Appendix: Extreme equilibrium phase angles in deceleration mode}\label{Appendix-unbound}
\begin{figure}[htb!]
    \centering
    \resizebox{1.0\linewidth}{!}
    {\includegraphics{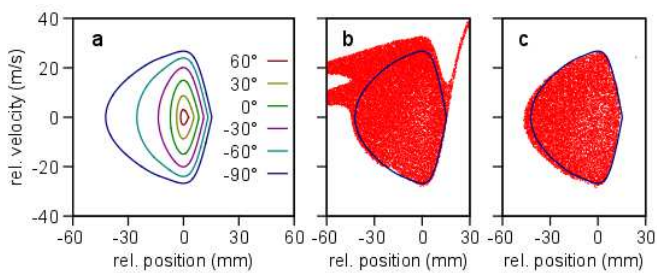}}
    \caption{Longitudinal phase-space diagrams showing the effect of negative $\phi_0$ in deceleration mode. (\emph{a}) shows the separatrices of this mode with different $\phi_0$, from 60$^{\circ}$ to -90$^{\circ}$ in steps of 30$^{\circ}$. (\emph{b,c}) show the results of 3D trajectory simulations of deceleration mode with $\phi_0 = -90^{\circ}$ after 100 and 200 deceleration stages, respectively. The graph is filled with dots that represent molecules of NH($X\,^3\Sigma^-, N=0, J=1$). These particles were initially distributed with a block function that well exceeded the separatrix.}
    \label{fig:unbound}
\end{figure}

As can be seen in Figure \ref{fig:acceptance-overview}\emph{b}, the highest acceptance is found with $\phi_0 = -90^{\circ}$ in deceleration mode or $\phi_0 = 90^{\circ}$ in acceleration mode. This is a surprising result if we consider conventional multi-stage Zeeman decelerators. In these decelerators, the inherent transverse defocusing fields outside the solenoids prevent the effective use of these extreme values of $\phi_0$\cite{Wiederkehr:JCP135:214202}. However, with the addition of magnetic hexapoles this limitation no longer exists. Indeed, the total acceptance changes almost solely with the longitudinal acceptance. This acceptance increases in deceleration and acceleration mode with lower and higher $\phi_0$, respectively. We show this for deceleration mode in Figure \ref{fig:unbound}\emph{a}.

In the negative $\phi_0$ range of deceleration mode the solenoids are turned off early, resulting in only a small amount of deceleration per stage. This is reflected in the kinetic energy change with this mode shown in Figure \ref{fig:acceptance-overview}\emph{a}. With lower $\phi_0$, less of the slope of the solenoid field is used to decelerate, and more of it is available for longitudinal focusing of the particle beam. 

Moreover, the minimum value of $\phi_0 = -90^{\circ}$ is an arbitrary limit, as even lower values of $\phi_0$ would produce even less deceleration and more longitudinal acceptance. Nevertheless, it is important to remember that this is a theoretical prediction, with the important assumption that the decelerator is of sufficient length that the slower particles have sufficient time to catch up with the synchronous particle. With less deceleration of the synchronous particle per solenoid, this catch-up time will increase. This is reflected in the difference in longitudinal phase-space distributions after 100 and 200 stages in Figures \ref{fig:unbound} (b) and (c), respectively. In these simulations (similar to those shown in Figure \ref{fig:phasespace3D}) a block distribution of NH($X\,^3\Sigma^-, N=0, J=1$) particles was used that well exceeded the predicted longitudinal separatrix. After 100 stages, the (deformed) corners of the initial block distribution are still visible as they revolve around the synchronous particle in longitudinal phase-space. Only after 200 stages of deceleration have these unaccepted particles had enough time to spatially separate from the particles with stable trajectories. This graph also shows that the prediction of the separatrix is quite accurate, and the uniformity of the particle distribution within is evidence of transverse phase stability, even with these extreme values of $\phi_0$. In acceleration mode, a similar rise in acceptance can be found with increasing $\phi_0$, which is also visible in Figure \ref{fig:acceptance-overview}\emph{b}.


\end{document}